RESEARCH ARTICLE

# Automated claustrum segmentation in human brain MRI using deep learning

Hongwei Li[1,2] | Aurore Menegaux[3,4] | Benita Schmitz-Koep[3,4] | Antonia Neubauer[3,4] | Felix J. B. Bäuerlein[5] | Suprosanna Shit[1] | Christian Sorg[3,4,6] | Bjoern Menze[1,2] | Dennis Hedderich[3,4]

[1]Department of Informatics, Technical University of Munich, Munich, Germany
[2]Department of Quantitative Biomedicine, University of Zurich, Zurich, Switzerland
[3]TUM-NIC Neuroimaging Center, Munich, Germany
[4]Department of Neuroradiology, Klinikum rechts der Isar, School of Medicine, Technical University of Munich, Munich, Germany
[5]Department of Molecular Structural Biology, Max Planck Institute of Biochemistry, Munich, Germany
[6]Department of Psychiatry, Klinikum rechts der Isar, School of Medicine, Technical University of Munich, Munich, Germany

**Correspondence**
Hongwei Li, IMETUM, Boltzmannstrasse 11, Garching, 85748, Bavaria, Germany.
Email: hongwei.li@tum.de

**Funding information**
German Federal Ministry of Education and Science, Grant/Award Numbers: BMBF 01ER0801, BMBF 01ER0803; Technische Universität München, Grant/Award Number: KKF 8765162; Federal Ministry of Education and Science, Grant/Award Numbers: BMBF 01ER0803, BMBF 01ER0801; Deutsche Forschungsgemeinschaft, Grant/Award Number: SO 1336/1-1

## Abstract

In the last two decades, neuroscience has produced intriguing evidence for a central role of the claustrum in mammalian forebrain structure and function. However, relatively few in vivo studies of the claustrum exist in humans. A reason for this may be the delicate and sheet-like structure of the claustrum lying between the insular cortex and the putamen, which makes it not amenable to conventional segmentation methods. Recently, Deep Learning (DL) based approaches have been successfully introduced for automated segmentation of complex, subcortical brain structures. In the following, we present a multi-view DL-based approach to segment the claustrum in T1-weighted MRI scans. We trained and evaluated the proposed method in 181 individuals, using bilateral manual claustrum annotations by an expert neuroradiologist as reference standard. Cross-validation experiments yielded median volumetric similarity, robust Hausdorff distance, and Dice score of 93.3%, 1.41 mm, and 71.8%, respectively, representing equal or superior segmentation performance compared to human intra-rater reliability. The leave-one-scanner-out evaluation showed good transferability of the algorithm to images from unseen scanners at slightly inferior performance. Furthermore, we found that DL-based claustrum segmentation benefits from multi-view information and requires a sample size of around 75 MRI scans in the training set. We conclude that the developed algorithm allows for robust automated claustrum segmentation and thus yields considerable potential for facilitating MRI-based research of the human claustrum. The software and models of our method are made publicly available.

KEYWORDS
claustrum, deep learning, image segmentation, MRI, multi-view

## 1 | INTRODUCTION

The claustrum is a highly conserved gray matter structure of the mammalian forebrain, situated in the white matter between the putamen and the insular cortex, more precisely between the external and the extreme capsule (Kowiański, Dziewiątkowski, Kowiańska, & Moryś, 1999; Puelles, 2014). Although first described by Félix Vicq d'Azyr in the late 18th century, it has remained one of the most enigmatic structures of the brain (Johnson & Fenske, 2014). In a seminal article by Sir Francis Crick and Christof Koch, they proposed a role of the claustrum for processes that give rise to integrated conscious percepts (Crick & Koch, 2005), which has spurred new interest in the claustrum and its putative function. From animal and human studies, we know that the claustrum is the most widely connected gray matter structure in the brain in relation to its size, being connected to both the ipsilateral and the contralateral hemisphere (Mathur, 2014; Pearson, Brodal, Gatter, & Powell, 1982; Reser et al., 2017; Torgerson, Irimia, Goh, & Van Horn, 2015; Zingg et al., 2014; Zingg, Dong, Tao, & Zhang, 2018). It is reciprocally connected to almost all cortical regions including motor and somatosensory as well as visual, limbic, auditory, associative, and prefrontal cortices, and receives neuromodulatory input from subcortical structures (Goll, Atlan, & Citri, 2015; Torgerson et al., 2015). While the claustrum's exact function remains elusive, recent evidence suggests a role in basic cognitive processes such as selective attention or task switching (Brown et al., 2017; Mathur, 2014; Remedios, Logothetis, & Kayser, 2010, 2014). A rather new but equally interesting perspective on the claustrum is its unique ontogeny and a link to so-called subplate neurons, which have been proposed to play a role in neurodevelopmental disorders such as schizophrenia, autism, and preterm birth (Bruguier et al., 2020; Hoerder-Suabedissen & Molnár, 2015; Watson & Puelles, 2017).

Most human in vivo studies using MRI to investigate the claustrum suffer from small sample sizes (Arrigo et al., 2017; Krimmel et al., 2019; Milardi et al., 2015) since the sheet-like and delicate anatomy of the claustrum precludes classic atlas-based segmentation methods and is challenging for statistical shape models and traditional machine learning methods (Aljabar, Wolz, & Rueckert, 2012; Heimann & Meinzer, 2009). Consequently, manual annotation has typically been necessary, which is notoriously time-consuming, requires expert knowledge, and is not feasible to be applied in large-scale studies of the human brain (Arrigo et al., 2017; Milardi et al., 2015; Torgerson & Van Horn, 2014).

Thus, to promote our understanding of the claustrum in humans, an objective and accurate, automated, MRI-based segmentation method is needed. As mentioned before, the claustrum is not included as a region of interest (ROI) in most MR-based anatomic atlases of the brain. In fact, only BrainSuite, which is a tool for automated cortical parcellation and subcortical segmentation based on surface-constrained volumetric registration of individual MR images of the brain to a manually labeled atlas, contains the claustrum as a ROI (Joshi, Shattuck, Thompson, & Leahy, 2007). However, this method has been shown to be rather unreliable, most likely due to the challenging anatomy of the claustrum (Berman, Schurr, Atlan, Citri, & Mezer, 2020). Very recently, an automated, rule-based method using anatomical landmarks for claustrum segmentation has been published and showed improved segmentation accuracy compared to BrainSuite but still only less accuracy in comparison with manual annotations (Berman et al., 2020). In conclusion, there is still the need for improved fast and reproducible, automated segmentation of the claustrum in order to enable its exploration in large MRI studies.

In recent years, computer vision and machine learning techniques have been increasingly used in the medical field, pushing the limits of segmentation methods relying on atlases, statistical shape models and traditional machine learning approaches (Aljabar et al., 2012; Aljabar, Heckemann, Hammers, Hajnal, & Rueckert, 2009; Heimann & Meinzer, 2009). Particularly, deep learning (DL) (LeCun, Bengio, & Hinton, 2015) based approaches have shown promising results on various medical image segmentation tasks, for example, brain structure and tumor segmentation in MR images (Chen, Dou, Yu, Qin, & Heng, 2018; Kamnitsas et al., 2017; Prados et al., 2017; Wachinger, Reuter, & Klein, 2018). Recent segmentation methods commonly rely on so-called convolutional neural networks (CNNs). Applied to segmentation tasks, these networks "learn" proper structural information from a set of manually labeled data serving as ground truth for training. In the testing stage, CNNs perform automated segmentation on unseen images yielding rather high accuracies even for tiny structures such as white-matter lesions (Li et al., 2018). Recently, a clustering-based approach was proposed to segment the dorsal claustrum (Berman et al., 2020) and achieved <60% Dice coefficient when compared with manual segmentations. Yet DL-based approaches, which leverage large-scale datasets, have not been explored and can potentially improve segmentation accuracy.

Thus, we hypothesized that DL-based techniques used to segment the claustrum can fill the existing gap. Based on a large number of manually annotated, T1-weighted brain MRI scans, we propose a 2D multi-view framework for fully automated claustrum segmentation. In order to assess our central hypothesis, we will evaluate the segmentation accuracy of our algorithm on an annotated dataset using three canonical evaluation metrics and compare it to intrarater variability. Further, we will investigate whether multi-view information significantly improves the segmentation performance. In addition, we will address the questions of robustness against scanner type and how increasing the training set affects segmentation accuracy. To foreshadow results, we found robust, reliable, and stable claustrum segmentation based on our DL algorithm, which we make publicly available using an open-source repository: https://github.com/hongweilibran/claustrum_multi_view.

## 2 | MATERIALS AND METHODS

### 2.1 | Datasets

In the following two sections, we describe the datasets and evaluation metrics used in this study. T1-weighted three-dimensional scans of 181 individuals without known brain injury were included from the Bavarian Longitudinal Study (Hedderich et al., 2019). The study was carried out following the *Declaration of Helsinki* and was approved by the local institutional review boards. Written consent was obtained from all participants. The MRI acquisition took place at two sites: The Department of Neuroradiology, Klinikum rechts der Isar, Technische Universität München ($n = 120$) and the Department of Radiology, University Hospital of Bonn ($n = 61$). MRI examinations were performed at both sites on either a *Philips Achieva 3 T* or a *Philips Ingenia 3 T* system using 8-channel SENSE head-coils.

The imaging protocol includes a high-resolution T1-weighted, 3D-MPRAGE sequence (TI = 1300 ms, TR = 7.7 ms, TE = 3.9 ms, flip angle 15°; field of view: 256 mm × 256 mm) with a reconstructed isotropic voxel size of 1 mm$^3$. All images were visually inspected for artifacts and gross brain lesions that could potentially impair manual claustrum segmentation (Table 1).

## 2.2 | Preprocessing

Before manual segmentation, the images are skull-stripped using *ROBEX* (Iglesias, Liu, Thompson, & Tu, 2011) and denoised with spatially adaptive nonlocal means (Manjón, Coupé, Martí-Bonmatí, Collins, & Robles, 2010) to enhance the visibility of the claustrum. Manual annotations were performed by a neuroradiologist (D.M.H.) with 7 years of experience using a modified segmentation protocol (Davis, 2008) in ITK-SNAP 3.6.0 (Yushkevich et al., 2006). In brief, the claustrum was segmented in axial and coronal orientations, including its dorsal and ventral division at individually defined optimal image contrast for differentiation of gray and white matter. First, the claustrum was delineated on axial slices at the basal ganglia level, where it is visible continuously. Second, the claustrum was traced inferiorly until it was no longer visible. Consecutively, the claustrum was traced superiorly until its superior border. Notably, the superior parts of the claustrum are usually discontinuous below the insular cortex. Then, the axial annotations were checked and corrected (if necessary) using coronal views. This process is essential for the claustrum parts extending below the putamen and the ventral claustrum extending to the stem of the temporal lobe.

An additional preprocessing step is performed on top of the basic preprocessing steps carried out by the rater. We aim to normalize the voxel intensities to reduce the variations across subjects and scanners. Thus, a simple yet effective preprocessing step is used in both training and testing stages. It includes two steps: (1) cropping or padding each slice to a uniform size and (2) *z-score* normalization of the brain voxel intensities. First, all the axial and coronal slices are automatically cropped or padded to 180 × 180 to guarantee a uniform input size for the deep-learning model. Next, z-score normalization is performed for individual 3D scans. The mean and standard deviation are calculated based on the intensities within each individual's brain mask. Finally, the voxel intensities are rescaled to a mean of zero and unit standard deviation (Figure 1).

## 2.3 | Multi-view fully convolutional neural networks

### 2.3.1 | Multi-view learning

When performing manual annotations, neuroradiologists rely on axial and coronal views to identify the structure. Thus, we hypothesized that the image features from the two geometric views would be complementary to locate the claustrum and would be beneficial for reducing false positives on individual views. We train two deep CNN models on 2D single-view slices after parsing a 3D MRI volume into axial and coronal views. The sagittal view is excluded because we find it does not improve segmentation results. Further discussion is provided in Section 3.2. We propose a practical and straightforward approach to aggregate the multi-view information in probability space at a voxel-wise level during the inference stage (see Figure 2a). We train two single-view models on the 2D image slices from axial and coronal views, respectively. During the testing stage, we predict the single-view segmentation mask and fuse the multi-view information by averaging the voxel-wise probabilities.

### 2.3.2 | Single-view 2D convolutional network architecture

We built a 2D architecture based on a recent U-shape network (Li et al., 2018; Ronneberger, Fischer, & Brox, 2015) and tailored it for the claustrum segmentation task. The network architecture is delineated in Figure 2. It consists of a down-convolutional part that shrinks the spatial dimensions (left side), and an up-convolutional part that expands the score maps (right side). Skip connections between down-convolutional and up-convolutional are used. In this model, two convolutional layers are repeatedly employed, followed by a rectified

**TABLE 1** Characteristics of the dataset in this study

| Datasets | Scanner name | Voxel size (mm$^3$) | Number of subjects |
|---|---|---|---|
| Bonn-1 | Philips Achieva 3 T | 1.00 × 1.00 × 1.00 | 15 |
| Bonn-2 | Philips Ingenia 3 T | 1.00 × 1.00 × 1.00 | 46 |
| Munich-1 | Philips Achieva 3 T | 1.00 × 1.00 × 1.00 | 103 |
| Munich-2 | Philips Ingenia 3 T | 1.00 × 1.00 × 1.00 | 17 |

*Note*: The dataset consists of 181 subjects from four scanners and two centers.

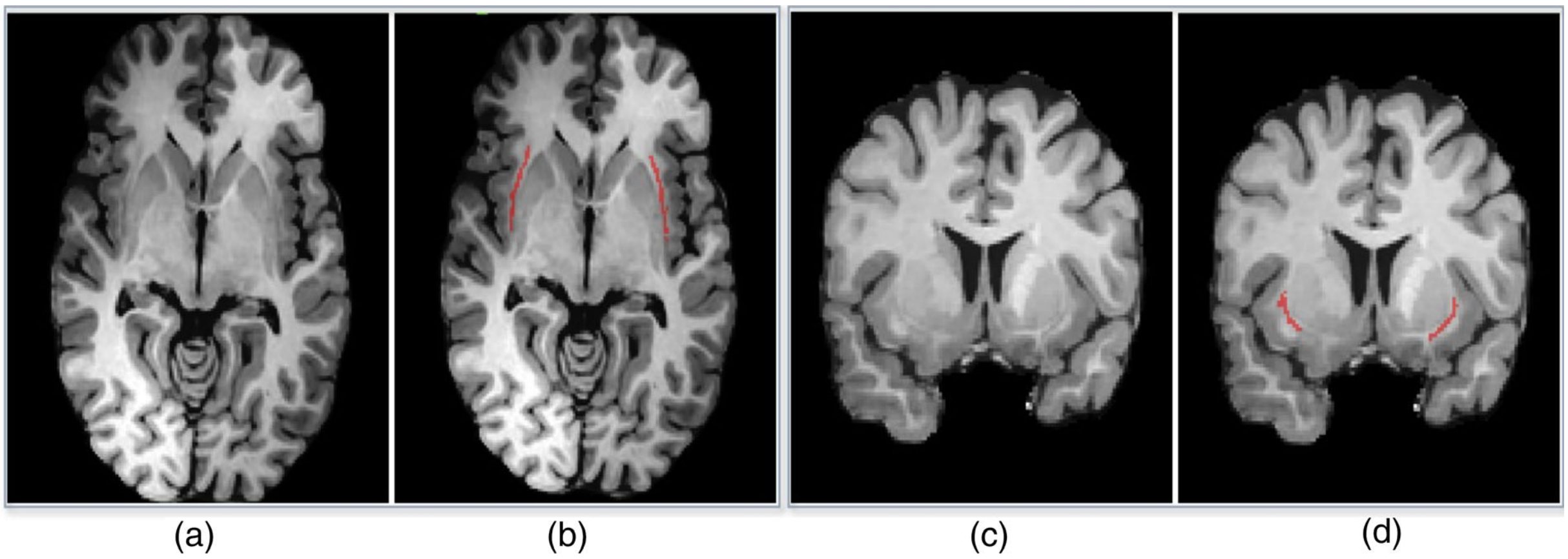

**FIGURE 1** Examples of axial (a, b) and coronal (c, d) MR slices with corresponding manual annotation of the claustrum structure (in b and d) by a neuroradiologist

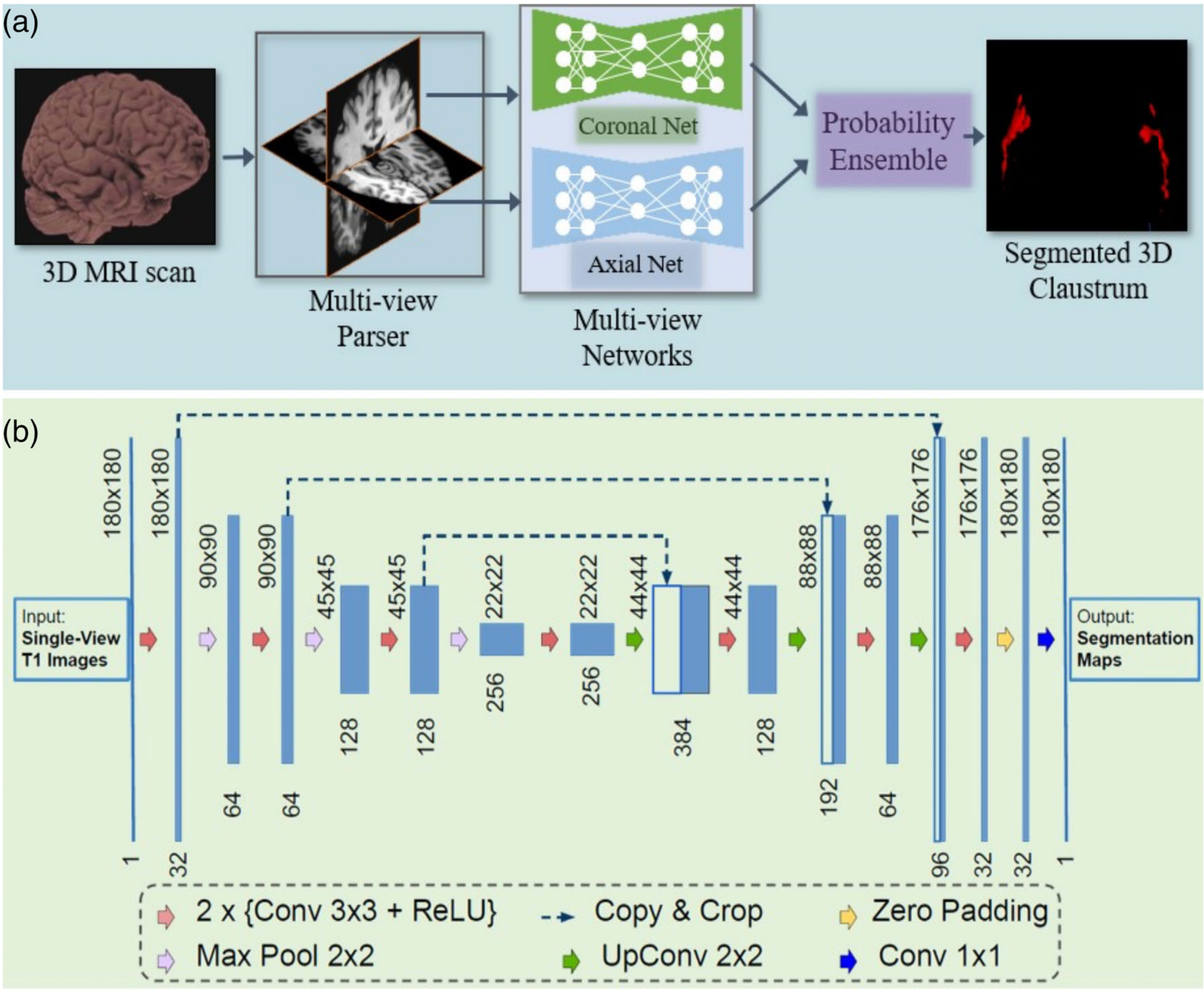

**FIGURE 2** (a) A schematic view of the proposed segmentation method using multi-view fully convolutional networks to segment the 3D claustrum jointly; (b) 2D Convolutional network architecture for each view (i.e., axial and coronal). It takes the raw images as input and predicts its segmentation maps. The network consists of several nonlinear computational layers in a shrinking part (left side) and an expansive part (right side) to extract semantic features of the claustrum structure linear unit (*ReLU*) and a $2 \times 2$ max pooling operation with stride 2 for down-sampling. At the final layer, a $1 \times 1$ convolution is used to map each 64-component feature vector to two classes. In total, the network contains 16 convolutional layers. The network takes the single-view slices of T1 modality scans as the input during training and testing (see Figure 2b).

## 2.4 | Loss function

With respect to the claustrum segmentation task, the numbers of positives (claustrum) and negatives (nonclaustrum) are highly unbalanced. One promising solution to tackle this issue is to use Dice loss (Milletari, Navab, & Ahmadi, 2016) as the loss function for training the model. The formulation is as follows.

Let $G=\{g_1,...,g_N\}$ be the ground-truth segmentation maps over $N$ slices, and $P=\{p_1,...,p_N\}$ be the predicted probabilistic maps over $N$ slices. The Dice loss function can be expressed as:

$$\text{Dice Loss} = -\frac{2\sum_{n=1}^{N}|p_n \circ g_n| + s}{\sum_{n=1}^{N}(|p_n|+|g_n|)+s}$$

where $\circ$ represents the entrywise product of two matrices, and $|\cdot|$ represents the sum of the matrix entries. The $s$ term is used here to ensure the loss function stability by avoiding the division by 0, that is, in a case where the entries of $G$ and $P$ are all zeros. $s$ is set to 1 in our experiments.

## 2.5 | Anatomically consistent postprocessing

The postprocessing for the 3D segmentation result included: (1) cropping or padding the segmentation maps concerning the original size, that is, an inverse operation to the step described in Section 2.3.1; (2) removing anatomically unreasonable artifacts. To remove unreasonable segmentations (e.g., the claustrum does not appear in the first and last slices which contain skull or other tissues), we employed a simple strategy: if there is a claustrum structure detected in the first $m$ and last $n$ ones of a brain along the z-direction, they are considered false positives. Empirically, m and n are set to 20% of the number of axial slices for each scan. The codes and models of the proposed method are made publicly available on *GitHub*.

## 2.6 | Parameter setting and computation complexity

An appropriate parameter setting is crucial to the successful training of deep convolutional neural networks. We selected the number of epochs to stop the training by contrasting training loss and the performance on validation set over epochs in each experiment, as shown in Figure S2 in the Supplement. Hence, we choose a number of $N$ epochs to avoid over-fitting and to keep a low computational cost by observing the VS and DSC on the validation set. The batch size was empirically set to 30 and the learning rate was set to 0.0002 throughout all experiments by observing the training stability on the validation set.

The experiments are conducted on a GNU/Linux server running Ubuntu 18.04, with 64GB RAM. The number of trainable parameters in the proposed model with one-channel inputs (T1) is 4,641,209. The algorithm was trained on a single NVIDIA Titan-V GPU with 12GB RAM. It takes around 100 min to train a single model for 200 epochs on a training set containing 5000 images with a size of 180 × 180 pixels. For testing, the segmentation of one scan with 192 slices by an ensemble of two models takes around 90 s using an Intel Xeon CPU (E3-1225v3) (without GPU use). In contrast, the segmentation per scan takes only 3 s when using a GPU.

## 2.7 | Evaluation metrics and protocol

Three metrics are used to evaluate the segmentation performance in different aspects in the reported experiments. For example, given a ground truth segmentation map $G$ and a predicted segmentation map $P$ generated by an algorithm, the evaluation metrics are defined as follows.

### 2.7.1 | Volumetric similarity (VS)

Let $V_G$ and $V_P$ be the volumes of region of interests in $G$ and $P$, respectively. Then the volumetric similarity (VS) in percentage is defined as:

$$VS = 1 - \frac{|V_G - V_P|}{|V_G + V_P|}$$

### 2.7.2 | Hausdorff distance (95th percentile) (HD95)

$$H(G,P) = \max\left\{\sup_{x \in G}\inf_{y \in P} d(x,y), \sup_{y \in G}\inf_{x \in P} d(x,y)\right\}$$

Where $d(x,y)$ denotes the distance of x and y, sup denotes the *supremum* and *inf* for the infimum. This measures the distance between the two subsets of metric space. It is modified to obtain a robust metric by using the 95th percentile instead of the maximum (100th percentile) distance.

### 2.7.3 | Dice similarity coefficient

$$DSC = \frac{2(G \cap P)}{|G| + |P|}$$

This measures the overlap between ground truth maps G and prediction maps P.

We use $k$-fold cross-validation to evaluate the overall performance. In each split, 80% scans from each scanner are pooled into a training set and the remaining scans as a test set. This procedure is repeated until all of the subjects were used in the testing phase.

## 3 | RESULTS

### 3.1 | Manual segmentation: intra-rater variability

In order to set a benchmark accuracy for manual segmentation, intra-rater variability was assessed based on repeated annotations of 20 left and right claustra by the same experienced neuroradiologist. In order to assure independent segmentation, annotations were performed at least 3 months apart. We obtained the intra-rater variability on 20 scans using the metrics VS, DSC, and HD95 and report the following median values with interquartile ranges (IQR): VS: 0.949, [0.928, 0.972]; DSC: 0.667, [0.642, 0.704], HD95: 2.24 mm, [2.0, 2.55]. Notably, the image resolution of all scans is 1.00 $mm^3$.

### 3.2 | DL-based segmentation: single-view vs. multi-view

In order to investigate the added value of multi-view information for the proposed system, we compare the segmentation performances of the single-view model (i.e., axial, coronal, or sagittal) with the multi-view ensemble model. To exclude the influence of scanner acquisition, we evaluate our method on the data from one scanner (*Munich-Achieva*), including 103 subjects and perform 5-fold cross-validation for a fair comparison. In each cross-validation split, the single-view CNNs and multi-view CNNs ensemble model are trained on images from the same subjects. Afterwards, they are evaluated on the test cases with respect to the evaluation metrics. Table 2 shows the segmentation performance of each setting. We observed that the sagittal view yields the worst performance among the three views (Figure 4).

We further perform statistical analysis (Wilcoxon signed-rank test) to compare the statistical significance between the proposed single-view CNNs and multi-view CNNs ensemble model. We observed that the two-view (*axial* + *coronal*) approach outperforms single-view ones significantly on HD95 and DSC. We further compared the three-view (axial, coronal, and sagittal) approach with the two-view(axial and coronal) approach and found that they are comparable in terms of VS, and that the two-view approach outperforms the three-view approach in terms of HD95 ($p = .035$) and DSC ($p = .021$). Thus, in the following sections, we use the *axial* + *coronal* two-view segmentation approach to evaluate the method.

### 3.3 | DL-based segmentation: stratified *k*-fold cross validation

In order to evaluate the general performance of our axial and coronal multi-view technique on the whole dataset, we performed stratified 5-fold cross validation. In each fold, we take 80% of subjects from all scanners, pool them into a training set and use the rest as a test set. Figure 3 and Table 3 show the segmentation performance of three metrics on 181 scans from four scanners, showing its effectiveness with respect to volume measurements and localization accuracy. In order to compare AI-based segmentation performance to the human expert rater benchmark performance, we performed *Wilcoxon signed-rank test* on 20 subjects as mentioned in Section 3.1 with respect to three evaluation metrics (see Table 3). We found no statistical difference between manual and AI-based segmentation with respect to VS, and we observed superior performance of AI-based segmentation with respect to HD95 and Dice score. This result indicates that AI-based segmentation performance is equal or superior to the human expert level.

### 3.4 | DL-based segmentation: Influence of individual scanners

To evaluate the generalizability of our method to unseen scanners, we present a leave-one-scanner-out study. For the cross-scanner analysis, we use the scanner IDs to split the 181 cases into training and test sets. In each split, the subjects from three scanners are used as a training set while the subjects from the remaining scanner are used as the test set. This procedure is repeated until all the scanners are used as test set. The achieved performance is comparable with the cross-

**TABLE 2** Segmentation performances (median values with IQR) of the single-view approaches and multi-view approaches

| | | | | | | *p* value | | |
|---|---|---|---|---|---|---|---|---|
| Metrics | Axial (A) | Coronal (C) | Sagittal (S) | A + C | A + C + S | A + C vs. A | A + C vs. C | A + C vs. A + C + S |
| VS (%) | 94.4 [90.1, 96.7] | 94.7 [90.4, 97.3] | 79.1 [73.5, 86.4] | 93.3 [89.6, 96.9] | 92.9 [89.6, 96.5] | .636 | **.008** | .231 |
| HD95↓ (mm) | 1.73 [1.41, 2.24] | 1.41 [1.41, 2.0] | 3.21 [2.24, 3.61] | 1.41 [1.41, 1.79] | 1.73 [1.41, 1.84] | **<.001** | **<.001** | **.035** |
| DSC (%) | 69.7 [66.0, 72.4] | 70.0 [67.2, 73.2] | 55.2 [45.7, 63.1] | 71.8 [68.7, 74.6] | 71.0 [68.5, 74.3] | <.001 | **<.001** | **.021** |

*Note*: Values in bold denote statistical significance. The combination of axial and coronal views shows its superiority over individual views. Note that we used equal weights for each view in the multi-view ensemble model.
Abbreviations: A, axial; C, coronal; DSC, dice similarity coefficient; HD95, 95th percentile of Hausdorff distance; S, sagittal; VS, volumetric similarity.

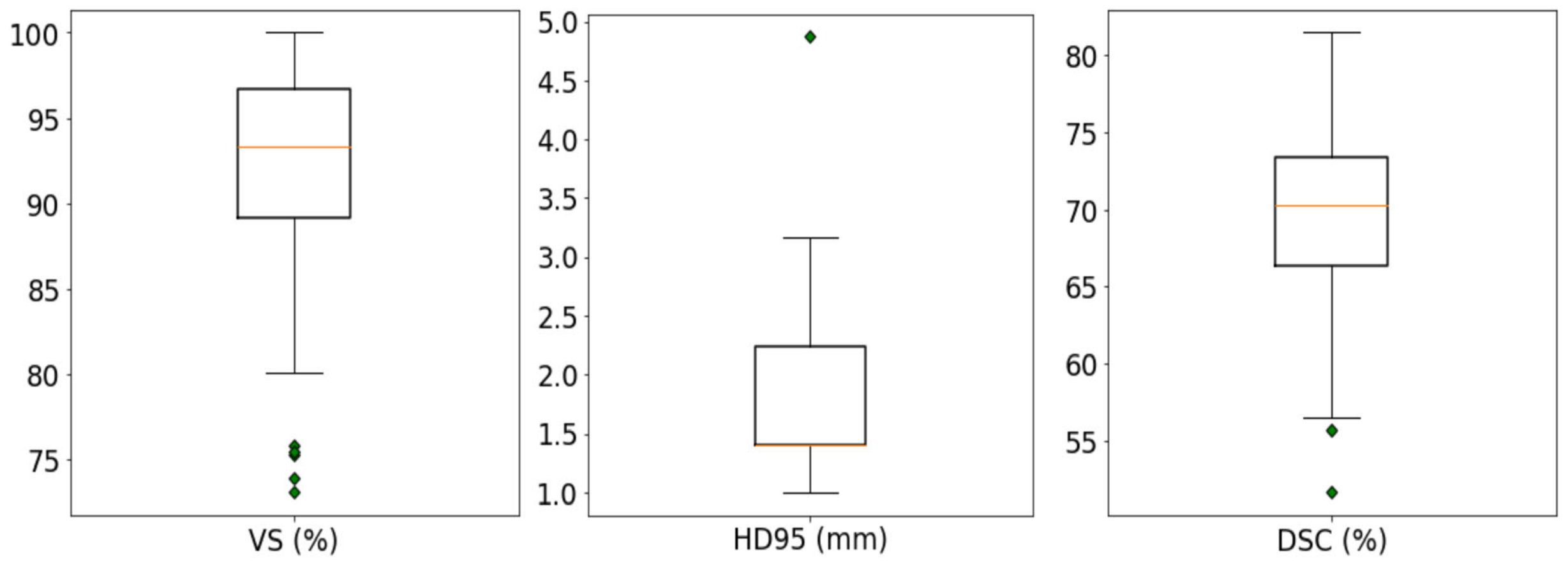

**FIGURE 3** Segmentation results of 5-fold cross-validation on the 181 scans across four scanners: *Bonn-Achieva*, *Bonn-Ingenia*, *Munich-Achieva*, and *Munich-Ingenia*. Each box plot summarizes the segmentation performance with respect to one specific evaluation metric

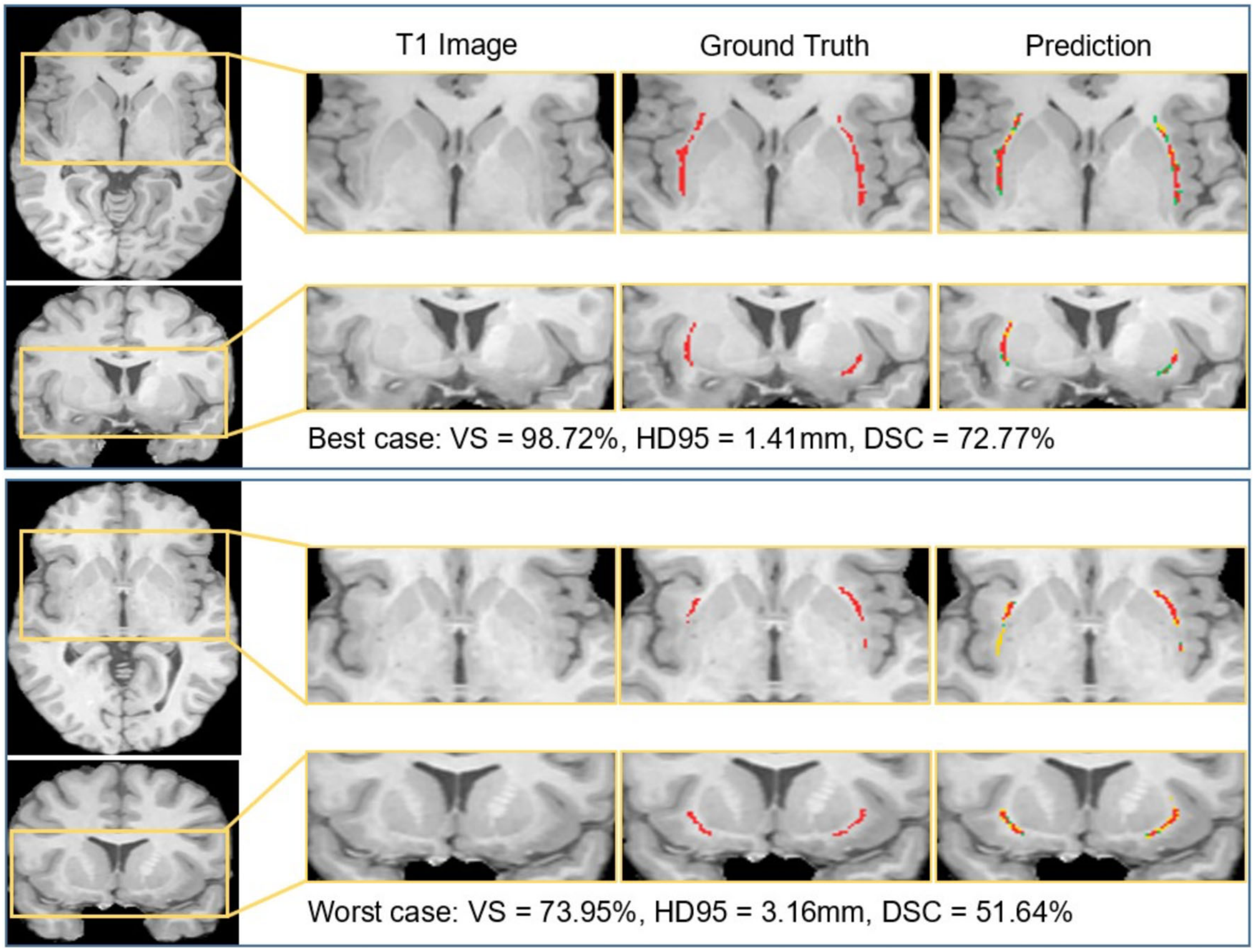

**FIGURE 4** Segmentation results of the best case and the worst case in terms of DSC. In the predicted segmentation masks, the red pixels represent true positives, the green ones represent false negatives, and the yellow ones represent false positives

validation results in Section 3.3, where all scanners were seen in the training set. Figure 5 plots the distributions of segmentation performances on four scanners being tested in turns. As shown in Table 4, we found that the cross-validation results achieved significantly lower HD95 and higher DSC than leave-one-scanner-out results at comparable VS. This is because for cross-validation, all scanners are included in training stage and thus no domain shift is seen between training and testing stages. This result indicates that testing the model on unseen scanners hampers segmentation performance.

To further investigate the influence of scanner acquisition for segmentation, we individually perform 5-fold cross-validation on the subsets *Bonn-Ingenia* and *Munich-Achieva* using subject IDs. The other two scanners are not evaluated because they contain relatively fewer scans. We use *Mann–Whitney U test* to compare the performance of

**TABLE 3** Performance comparison of manual and AI-based segmentations on 20 subjects with Wilcoxon signed-rank test

| Metrics | Manual segmentation [median, IQR] | AI-based segmentation [median, IQR] | *p* value |
|---|---|---|---|
| VS (%) | 94.9, [91.4, 97.6] | 94.3, [89.6, 96.7] | .821 |
| HD95 (mm) | 2.24, [2.0, 2.55] | 1.41, [1.41, 2.24] | **.005** |
| DSC (%) | 68.9, [64.2, 70.9] | 71.7, [67.8, 73.5] | **.001** |

*Note*: We found that AI-based segmentation performance is equal or superior to the human expert level.
Abbreviations: DSC, Dice similarity coefficient; HD95, 95th percentile of Hausdorff Distance; VS, volumetric similarity.

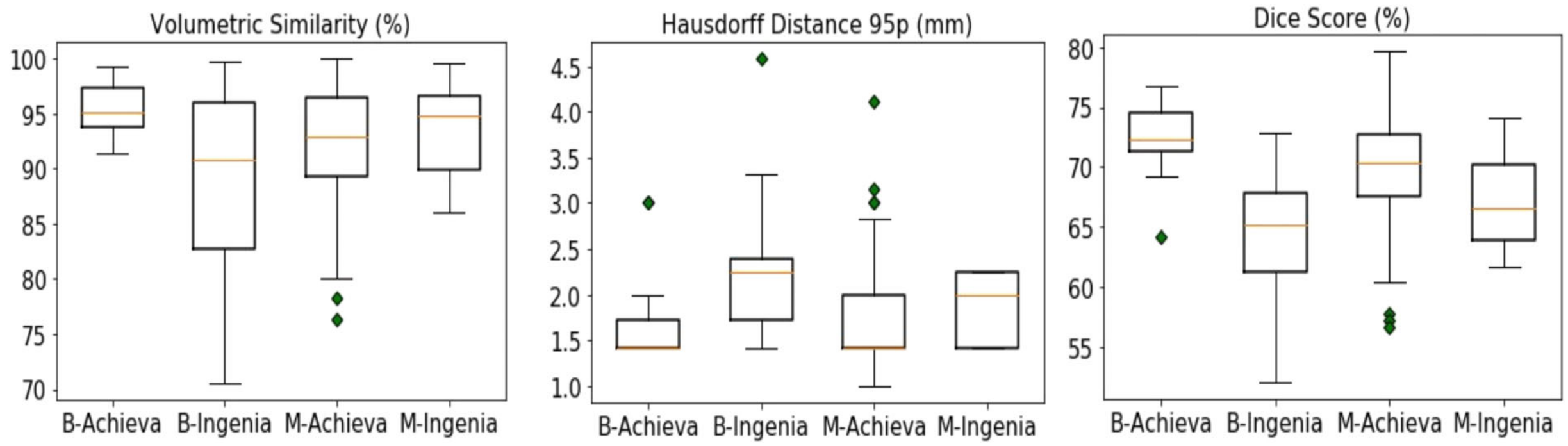

**FIGURE 5** Segmentation results of leave-one-scanner-out evaluation on the four scanners. Each sub-figure summarizes the segmentation performance on the testing scans from four scanners with respect to one metric. For example, the boxplot named *Bonn-Achieva* in the left sub-figure shows the distribution of segmentation results on scanner *Bonn-Achieva* (scanner 1) when using data from the other three scanners to train the AI model

**TABLE 4** Statistics analysis of leave-one-scanner-out segmentation results and *k*-fold cross-validation results

| Metrics | Leave-one-scanner-out (mean ± Std) | *k*-fold cross-validation (mean ± Std) | *p* value |
|---|---|---|---|
| VS (%) | 91.9 ± 6.2 | 92.2 ± 5.7 | .268 |
| HD95(mm)↓ | 1.86 ± 0.58 | 1.76 ± 0.51 | **<.001** |
| DSC (%) | 68.3 ± 5.0 | 69.5 ± 5.3 | **<.001** |

*Note*: Values in bold denote statistical significance. Statistical differences between them with respect to HD95 and Dice score were observed. It indicated that testing on unseen scanners harms the segmentation performance.
Abbreviations: DSC, Dice similarity coefficient; HD95, 95th percentile of Hausdorff Distance; VS, volumetric similarity.

the two groups. We found that *Bonn-Ingenia* obtained significantly lower VS and lower DSC than *Munich-Achieva*, which indicates that scanner characteristics such as image contrast, noise level, etc., generally affect the performance of AI-based segmentation. The box plots of the two evaluations are shown in Figure S1.

## 3.5 | How much training data is needed?

Since supervised deep learning is a data-driven machine learning method, it commonly requires a large amount of training data to optimize the nonlinear computational model. However, it is necessary to know the boundary when the model begins to saturate in order to avoid unnecessary manual annotations. Here, we perform a quantitative analysis on the effect of the amount of training data. Specifically, we split the 181 scans into a training set and a validation set with a ratio of 4:1 in a stratified manner from four scanners, resulting in 146 subjects for training and 35 for validation. As a start, we randomly pick 10% of the scans from the training set, train, and test the model. Then we gradually increased the size of the training set by a step of 10%. Figure 6 shows that the HD95 and the DSC only marginally improve on the validation set when >50% of the training set is used, while the VS is relatively stable over the whole range. Thus, we conclude that a training set including around 75 manually annotated scans is sufficient to obtain good segmentation results.

# 4 | DISCUSSION AND CONCLUSION

We have presented a deep-learning-based approach to accurately segment the claustrum, a complex and tiny gray matter structure of the human forebrain that has not been amenable to conventional segmentation methods. The proposed method uses multi-view information from T1-weighted MRI and achieves expert-level segmentation in a fully automated manner. To the best of our knowledge, this is the first work on fully automated segmentation of human claustrum using state-of-the-art deep learning techniques.

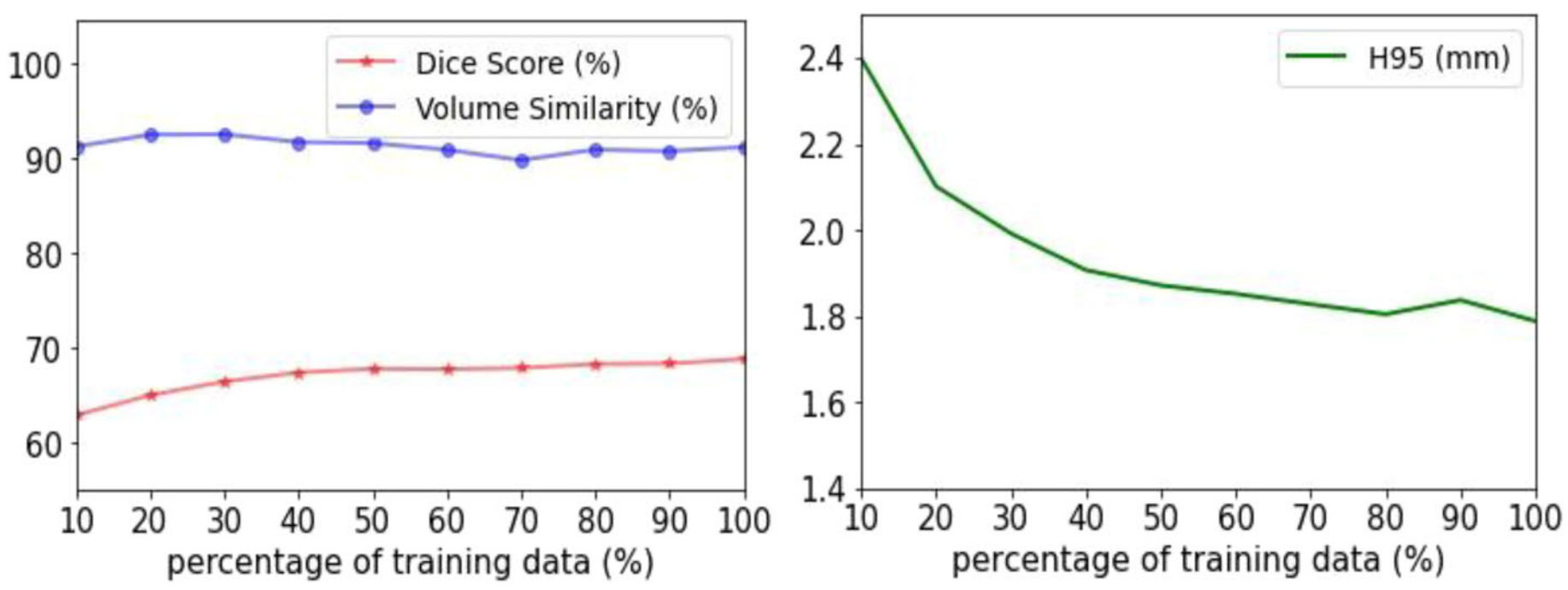

**FIGURE 6** Segmentation performance on the validation set when gradually increasing the percentage of the training data by a step of 10%. Only a marginal improvement on the validation set was observed when >50% of the training set was used

The first finding is that the segmentation performance benefits from leveraging multi-view information, specifically combining axial and coronal orientations. The significance of improvement was confirmed using paired difference tests. The multi-view fusion process imitates the annotation workflow by neuroradiologists, relying on 3D anatomical knowledge from multiple views. This strategy is also shown to be effective in common brain structure segmentation (Wei, Xia, & Zhang, 2019; Zhao, Zhang, Song, & Liu, 2019) and cardiac image segmentation (Mortazi et al., 2017). We observed that integrating sagittal view does not further improve the performance. This is because the claustrum, a thin, sheet-like structure is mainly oriented in the sagittal plane and can hardly be delineated in the sagittal view.

The proposed method yields a high median volumetric similarity, a small Hausdorff distance, and a decent Dice score in the cross-validation experiments. Although the achieved Dice score presents a relatively small value (~70%), we claim that this is excellent considering the structure of the claustrum is very tiny (usually <1500 voxels at 1 $mm^3$ isotropic resolution). We illustrate the correlation between Dice scores and claustrum volumes in the Supplement. In similar tasks such as segmentation of multiple sclerosis lesions with thousands of voxels, a Dice score of around 75% would be considered excellent. For the segmentation of larger tissues such as white matter and gray matter, Dice scores would reach 95% (Gabr et al., 2020). Nevertheless, HD95 quantifies the distance between prediction and ground-truth masks and is robust to assess tiny and thin structures (Kuijf et al., 2019).

Another valuable finding is that the proposed algorithm achieves expert-level segmentation performance and even outperforms a human expert rater in terms of DSC and HD95, which is confirmed by comparing the two groups of segmentation performances done by human rater and the proposed method. We conclude that the human rater presents more bias when the structure is tiny and ambiguous. Meanwhile, an AI-based algorithm learns to fit the available knowledge and shows a stable behavior when performing automated segmentation. This finding aligns recent advances in biomedical research where deep learning-based methods demonstrate unbiased quantification of structures (Todorov et al., 2020). Thus, we conclude that the proposed method would quantify the claustrum structure in an accurate and unbiased way.

We found that the segmentation performance slightly dropped when the AI-based model was tested on unseen scanners. Domain shift is commonly observed in machine learning tasks between training and testing data with different distributions. However, from our observation, the performance drop in the experiment is not severe, and the segmentation outcome is satisfactory. This is because scanners are in similar resolution from the same manufacturer, and the scans are properly pre-processed, resulting in a small domain gap. To enforce our model to be generalized to unseen scanners, domain adaptation methods (Dou, de Castro, Kamnitsas, & Glocker, 2019; Kamnitsas et al., 2017) are to be investigated in future studies.

Although the proposed method reaches expert-level performance and provides unbiased quantification results, our work has a few limitations. First, the human claustrum has a thin and sheet-like structure. Thus, high-resolution imaging as used in this study at an isotropic resolution of 1 $mm^3$ will result in partial volume effects, which significantly affect both the manual expert annotation and the automated segmentation. We addressed this bias by using a clear segmentation protocol to reduce variability in manual annotations as the reference standard. Second, the data distribution of the four datasets is highly imbalanced. It potentially affects the accuracy of the leave-one-scanner-out experiment in Section 3.4, especially when a significant sub-set (e.g., Munich-2) was taken out as a test set. In future work, evaluating the scanner influence on a more balanced dataset would avoid such an effect.

In conclusion, we described a multi-view deep learning approach for automatic segmentation of human claustrum structure. We empirically studied the effectiveness of multi-view information, leave-one-scanner-out study, the influence of imaging protocols and the effect of the amount of training data. We found that: (1) multi-view information, including coronal and axial views, provide complementary information to identify the claustrum structure; (2) multi-view automatic segmentation is equal or superior to manual segmentation accuracy; (3) scanner type affects segmentation accuracy even for identical sequence parameter settings; (4) a training set with 75 scans and annotation is sufficient to achieve satisfactory segmentation result. We have made our Python implementation codes available on *GitHub*.

## ACKNOWLEDGMENTS

We thank all current and former members of the Bavarian Longitudinal Study Group who contributed to general study organization, recruitment and data collection, management and subsequent analyses, including (in alphabetical order): Barbara Busch, Stephan Czeschka, Claudia Grünzinger, Christian Koch, Diana Kurze, Sonja Perk, Andrea Schreier, Antje Strasser, Julia Trummer, and Eva van Rossum. We are grateful to the staff of the Department of Neuroradiology in Munich and the Department of Radiology in Bonn for their help in data collection. Most importantly, we thank all our study participants and their families for their efforts to take part in this study. This study is supported by the Deutsche Forschungsgemeinschaft (SO 1336/1-1 to Christian Sorg), German Federal Ministry of Education and Science (BMBF 01ER0803 to Christian Sorg) and the Kommission für Klinische Forschung, Technische Universität München (KKF 8765162 to Christian Sorg). Open Access funding enabled and organized by Projekt DEAL.

## CONFLICT OF INTEREST

The authors declare no conflict of interest.

## DATA AVAILABILITY STATEMENT

The data that support the findings of this study are available from the corresponding author upon reasonable request.

## ORCID

*Hongwei Li* https://orcid.org/0000-0002-5328-6407
*Aurore Menegaux* https://orcid.org/0000-0003-3965-0396
*Benita Schmitz-Koep* https://orcid.org/0000-0002-8874-5749
*Dennis Hedderich* https://orcid.org/0000-0001-8994-5593

## REFERENCES

Aljabar, P., Heckemann, R. A., Hammers, A., Hajnal, J. V., & Rueckert, D. (2009). Multi-atlas based segmentation of brain images: Atlas selection and its effect on accuracy. *NeuroImage*, *46*(3), 726–738.

Aljabar, P., Wolz, R., & Rueckert, D. (2012). Manifold learning for medical image registration, segmentation, and classification. *Machine Learning in Computer-Aided Diagnosis: Medical Imaging Intelligence and Analysis*, *1*, 351–372.

Arrigo, A., Mormina, E., Calamuneri, A., Gaeta, M., Granata, F., Marino, S., … Quartarone, A. (2017). Inter-hemispheric claustral connections in human brain: A constrained spherical deconvolution-based study. *Clinical Neuroradiology*, *27*(3), 275–281.

Berman, S., Schurr, R., Atlan, G., Citri, A., & Mezer, A. A. (2020). Automatic segmentation of the dorsal claustrum in humans using in vivo high-resolution MRI. *Cerebral Cortex Communications*, *1*(1), 1–14. https://doi.org/10.1093/texcom/tgaa062

Brown, S. P., Mathur, B. N., Olsen, S. R., Luppi, P.-H., Bickford, M. E., & Citri, A. (2017). New breakthroughs in understanding the role of functional interactions between the neocortex and the claustrum. *Journal of Neuroscience*, *37*(45), 10877–10881.

Bruguier, H., Suarez, R., Manger, P., Hoerder-Suabedissen, A., Shelton, A. M., Oliver, D. K., … Puelles, L. (2020). In search of common developmental and evolutionary origin of the claustrum and subplate. *Journal of Comparative Neurology*, *528*(17), 2956–2977.

Chen, H., Dou, Q., Yu, L., Qin, J., & Heng, P.-A. (2018). VoxResNet: Deep voxelwise residual networks for brain segmentation from 3D MR images. *NeuroImage*, *170*, 446–455.

Crick, F. C., & Koch, C. (2005). What is the function of the claustrum? *Philosophical Transactions of the Royal Society B: Biological Sciences*, *360*(1458), 1271–1279.

Davis, W. B. (2008). The claustrum in autism and typically developing male children: a quantitative MRI study. Brigham Young University.

Dou, Q., de Castro, D. C., Kamnitsas, K. & Glocker, B. (2019). *Domain generalization via model-agnostic learning of semantic features*. Paper presented at the Advances in Neural Information Processing Systems 32, Vancouver, Canada, 6450-6461.

Gabr, R. E., Coronado, I., Robinson, M., Sujit, S. J., Datta, S., Sun, X., … Narayana, P. A. (2020). Brain and lesion segmentation in multiple sclerosis using fully convolutional neural networks: A large-scale study. *Multiple Sclerosis Journal*, *26*(10), 1217–1226.

Goll, Y., Atlan, G., & Citri, A. (2015). Attention: The claustrum. *Trends in Neurosciences*, *38*(8), 486–495.

Hedderich, D. M., Bäuml, J. G., Berndt, M. T., Menegaux, A., Scheef, L., Daamen, M., … Wolke, D. (2019). Aberrant gyrification contributes to the link between gestational age and adult IQ after premature birth. *Brain*, *142*(5), 1255–1269.

Heimann, T., & Meinzer, H.-P. (2009). Statistical shape models for 3D medical image segmentation: A review. *Medical Image Analysis*, *13*(4), 543–563.

Hoerder-Suabedissen, A., & Molnár, Z. (2015). Development, evolution and pathology of neocortical subplate neurons. *Nature Reviews Neuroscience*, *16*(3), 133–146.

Iglesias, J. E., Liu, C.-Y., Thompson, P. M., & Tu, Z. (2011). Robust brain extraction across datasets and comparison with publicly available methods. *IEEE Transactions on Medical Imaging*, *30*(9), 1617–1634.

Johnson, J. I., & Fenske, B. A. (2014). History of the study and nomenclature of the claustrum. In *The Claustrum* (pp. 1-27). Amsterdam: Elsevier.

Joshi, A. A., Shattuck, D. W., Thompson, P. M., & Leahy, R. M. (2007). Surface-constrained volumetric brain registration using harmonic mappings. *IEEE Transactions on Medical Imaging*, *26*(12), 1657–1669.

Kamnitsas, K., Baumgartner, C., Ledig, C., Newcombe, V., Simpson, J., Kane, A., Rueckert, D. & Glocker, B. (2017). *Unsupervised domain adaptation in brain lesion segmentation with adversarial networks*. Paper presented at the International Conference on Information Processing in Medical Imaging (pp. 597-609). Springer, Cham.

Kamnitsas, K., Ledig, C., Newcombe, V. F., Simpson, J. P., Kane, A. D., Menon, D. K., … Glocker, B. (2017). Efficient multi-scale 3D CNN with fully connected CRF for accurate brain lesion segmentation. *Medical Image Analysis*, *36*, 61–78.

Kowiański, P., Dziewiątkowski, J., Kowiańska, J., & Moryś, J. (1999). Comparative anatomy of the claustrum in selected species: A morphometric analysis. *Brain, Behavior and Evolution*, *53*(1), 44–54.

Krimmel, S. R., White, M. G., Panicker, M. H., Barrett, F. S., Mathur, B. N., & Seminowicz, D. A. (2019). Resting state functional connectivity and cognitive task-related activation of the human claustrum. *NeuroImage*, *196*, 59–67.

Kuijf, H. J., Biesbroek, J. M., De Bresser, J., Heinen, R., Andermatt, S., Bento, M., … Casamitjana, A. (2019). Standardized assessment of automatic segmentation of white matter hyperintensities and results of the WMH segmentation challenge. *IEEE Transactions on Medical Imaging*, *38*(11), 2556–2568.

LeCun, Y., Bengio, Y., & Hinton, G. (2015). Deep learning. *Nature*, *521*(7553), 436–444.

Li, H., Jiang, G., Zhang, J., Wang, R., Wang, Z., Zheng, W.-S., & Menze, B. (2018). Fully convolutional network ensembles for white matter hyperintensities segmentation in MR images. *NeuroImage*, *183*, 650–665.

Manjón, J. V., Coupé, P., Martí-Bonmatí, L., Collins, D. L., & Robles, M. (2010). Adaptive non-local means denoising of MR images with spatially varying noise levels. *Journal of Magnetic Resonance Imaging*, *31*(1), 192–203.

Mathur, B. N. (2014). The claustrum in review. *Frontiers in Systems Neuroscience*, *8*, 48.

Milardi, D., Bramanti, P., Milazzo, C., Finocchio, G., Arrigo, A., Santoro, G., … Gaeta, M. (2015). Cortical and subcortical connections of the human claustrum revealed in vivo by constrained spherical deconvolution tractography. *Cerebral Cortex*, *25*(2), 406–414.

Milletari, F., Navab, N. & Ahmadi, S.-A. (2016). *V-net: Fully convolutional neural networks for volumetric medical image segmentation*. Paper presented at the 2016 Fourth International Conference on 3D vision (3DV) (pp. 565-571). Piscataway, NJ:IEEE.

Mortazi, A., Karim, R., Rhode, K., Burt, J., & Bagci, U. CardiacNET: *Segmentation of left atrium and proximal pulmonary veins from MRI using multi-view CNN*. Paper presented at the International Conference on Medical Image Computing and Computer-Assisted Intervention. Cham,Switzerland: Springer.

Pearson, R., Brodal, P., Gatter, K., & Powell, T. (1982). The organization of the connections between the cortex and the claustrum in the monkey. *Brain Research*, *234*(2), 435–441.

Prados, F., Ashburner, J., Blaiotta, C., Brosch, T., Carballido-Gamio, J., Cardoso, M. J., … De Leener, B. (2017). Spinal cord grey matter segmentation challenge. *NeuroImage*, *152*, 312–329.

Puelles, L. (2014). Development and evolution of the claustrum. In *The Claustrum* (pp. 119–176). Amsterdam, The Netherlands: Elsevier.

Remedios, R., Logothetis, N. K., & Kayser, C. (2010). Unimodal responses prevail within the multisensory claustrum. *Journal of Neuroscience*, *30*(39), 12902–12907.

Remedios, R., Logothetis, N. K., & Kayser, C. (2014). A role of the claustrum in auditory scene analysis by reflecting sensory change. *Frontiers in Systems Neuroscience*, *8*, 44.

Reser, D. H., Majka, P., Snell, S., Chan, J. M., Watkins, K., Worthy, K., … Rosa, M. G. (2017). Topography of claustrum and insula projections to medial prefrontal and anterior cingulate cortices of the common marmoset (*Callithrix jacchus*). *Journal of Comparative Neurology*, *525*(6), 1421–1441.

Ronneberger, O., Fischer, P., & Brox, T. (2015). *U-net: Convolutional networks for biomedical image segmentation*. Paper presented at the International Conference on Medical Image Computing and Computer-Assisted Intervention. (pp. 234–241). Cham, Switzerland: Springer.

Todorov, M. I., Paetzold, J. C., Schoppe, O., Tetteh, G., Efremov, V., Völgyi, K., … Menze, B. (2020). Automated analysis of whole brain vasculature using machine learning. *Nature Methods*, *17*(4), 442–449.

Torgerson, C. M., Irimia, A., Goh, S. M., & Van Horn, J. D. (2015). The DTI connectivity of the human claustrum. *Human Brain Mapping*, *36*(3), 827–838.

Torgerson, C. M., & Van Horn, J. D. (2014). A case study in connectomics: The history, mapping, and connectivity of the claustrum. *Frontiers in Neuroinformatics*, *8*, 83.

Wachinger, C., Reuter, M., & Klein, T. (2018). DeepNAT: Deep convolutional neural network for segmenting neuroanatomy. *NeuroImage*, *170*, 434–445.

Watson, C., & Puelles, L. (2017). Developmental gene expression in the mouse clarifies the organization of the claustrum and related endopiriform nuclei. *Journal of Comparative Neurology*, *525*(6), 1499–1508.

Wei, J., Xia, Y., & Zhang, Y. (2019). M3Net: A multi-model, multi-size, and multi-view deep neural network for brain magnetic resonance image segmentation. *Pattern Recognition*, *91*, 366–378.

Yushkevich, P. A., Piven, J., Hazlett, H. C., Smith, R. G., Ho, S., Gee, J. C., & Gerig, G. (2006). User-guided 3D active contour segmentation of anatomical structures: Significantly improved efficiency and reliability. *NeuroImage*, *31*(3), 1116–1128.

Zhao, Y.-X., Zhang, Y.-M., Song, M. & Liu, C.-L. (2019). *Multi-view semi-supervised 3d whole brain segmentation with a self-ensemble network*. Paper presented at the International Conference on Medical Image Computing and Computer-Assisted Intervention (pp. 256-265). Springer, Cham.

Zingg, B., Dong, H. W., Tao, H. W., & Zhang, L. I. (2018). Input–output organization of the mouse claustrum. *Journal of Comparative Neurology*, *526*(15), 2428–2443.

Zingg, B., Hintiryan, H., Gou, L., Song, M. Y., Bay, M., Bienkowski, M. S., … Toga, A. W. (2014). Neural networks of the mouse neocortex. *Cell*, *156*(5), 1096–1111.

## SUPPORTING INFORMATION

Additional supporting information may be found in the online version of the article at the publisher's website.